\documentclass[12pt,manuscript,psfig,epsfig,pdffig,superscriptaddress,floatfix]{revtex4}

\usepackage[latin9]{inputenc}
\usepackage{amsmath}
\usepackage{amssymb}
\usepackage{graphicx}
\usepackage{epstopdf}
\usepackage{pstricks}
\usepackage{color}

\makeatletter
\@ifundefined{textcolor}{}
{%
	\definecolor{BLACK}{gray}{0}
	\definecolor{WHITE}{gray}{1}
	\definecolor{RED}{rgb}{1,0,0}
	\definecolor{GREEN}{rgb}{0,1,0}
	\definecolor{BLUE}{rgb}{0,0,1}
	\definecolor{CYAN}{cmyk}{1,0,0,0}
	\definecolor{MAGENTA}{cmyk}{0,1,0,0}
	\definecolor{YELLOW}{cmyk}{0,0,1,0}
}

\usepackage{dcolumn}% Align table columns on decimal point
\usepackage{bm}% bold math
\usepackage{amsfonts}
\usepackage{cancel}

\makeatother

\begin{document}

\title{Colour and baryon number fluctuation of  preconfinement system in  production process and   $T_{cc}$ structure}

\author{Yi Jin}
\affiliation{School of Physics and Technology, University of Jinan, Jinan 250022,
P. R. China}
\author{Shi-Yuan Li}
\affiliation{School of Physics, Shandong University, Jinan 250100, P. R. China}
\author{Yan-Rui Liu}
\affiliation{School of Physics, Shandong University, Jinan 250100, P. R. China}

\author{Qin Qin}
\affiliation{School of physics, Huazhong University of Science and Technology, Wuhan 430074, China}

\author{Zong-Guo Si}
\affiliation{School of Physics, Shandong University, Jinan 250100, P. R. China}

\author{Fu-Sheng Yu}
\affiliation{School of Nuclear Science and Technology,  and Frontiers Science Center for Rare Isotopes, Lanzhou University, Lanzhou 730000, China}
\affiliation{Lanzhou Center for Theoretical Physics, and Key Laboratory of Theoretical Physics of Gansu Province, Lanzhou University, Lanzhou 730000, China}
\affiliation{Center for High Energy Physics, Peking University, Beijing 100871, China}

\begin{abstract}
We suggest to study the production mechanism and some details of the production properties to probe the structure of
the $DD\pi$ resonance $T_{cc}^+$ recently observed by the LHCb Collaboration. If the resonance is produced as a four-quark state,
one can find the corresponding finger prints via measurements on some production properties that are the same as those of $\Xi_{cc}$.
On the other hand, if $T_{cc}^+$ is produced as a hadron molecule, the measurement on the momentum correlation of $DD^*$ can be a smoking gun to make the judgement.
\end{abstract}

 \date{\today}
\maketitle	

 \section{Introduction}

The quark model predicts quark-antiquark and three-quark hadrons, i.e., mesons and baryons.  In these hadrons,  quarks are bound by the strong interactions and they are in colour singlet. However, a cluster of more (anti)quarks can also be in colour singlet. It is possible for this kind of colour-singlet clusters with modest mass to be a 'real' particle.  These belong to  the exotic hadrons, among whom the four-quark states ($qq\bar q\bar q$, tetraquark) and five-quark states ($qqqq\bar q$, pentaquark) with at least one heavy quark ($c,~b$) have been observed.  %\textcolor{blue}{ {\large 1} which review paper(s) ?some above statements need some refs.??}
 % Many exotic have been observed ..studied.. However, they
Almost all the four- and five-quark states are observed to be produced from decays of heavier hadrons (e.g., bottom hadrons) rather than promptly produced from multi-production processes in experiments.  From the theoretical aspects, this fact is understood that their production rate generally is quite small in multiproduction processes because of the unitarity constraint \cite{Han:2009jw, Jin:2016vjn} as well as the modest mass of the preconfinement clusters \cite{Li:2019vrc}.

All kinds of the multiquark state hadrons have one common property that
the bound (anti)quarks inside can be grouped into several clusters, with each
cluster {\it possibly} in colour-singlet. Hence, these hadrons could just
be hadron molecules rather than real multiquark states. However, the ways of grouping
these (anti)quarks  in a  multiquark state hadron are not unique, as it is simply known from the group
theory that the reduction ways for a direct product of several
representations are not unique. Furthermore, these clusters in one hadron need
not necessarily be in colour-singlet, since the only requirement is that the
whole set of these clusters are in colour-singlet. For example, the system $q_1
\bar{q_2} q_3 \bar{q_4}$ (the constituents of a four quark state) can
be decomposed/clustered in the following ways:
\begin{eqnarray}
(q_1 q_3)_{\bar 3} \otimes (\bar{q_2}
\bar{q_4})_{3} \rightarrow 1\\
(q_1 \bar q_2)_{1~ or ~8} \otimes (q_3
\bar q_4)_{1~ or ~8} \rightarrow 1\\
\cdot \cdot \cdot \nonumber
\end{eqnarray}
Here we just mention that such group theory analysis is applicable to the
quark states  as well as the quark field operators \cite{pire}.
In the above example, only  the second case, when these two $q\bar
q$ pairs are both in colour-singlet, it seems possible to be considered
as a hadron molecule. However dynamically, the colour interactions in
the system via exchanging gluons can change the colour state of each
individual  cluster, so each kind of grouping/reduction way seems having no special physical
reasons. Such an ambiguity, which has been considered in many
 hadronization and decay processes as ``colour
recombination/rearrangement'' \cite{our1, our2, gsj}, obstacles the possibility
to consider the multiquark hadron in a unique and uniform way, while
leads to the possibility of introducing some phenomenological
duality. Namely, even we consider the production of multiquark hadron as
``hadron molecule'' formation, the subsequent colour interactions
 in the system can eventually transit  this
``molecule'' into a ``real''  multiquark hadron, at least by some
probability --- et vice verse.

However,  this ambiguity just gives us a clue: even though we can not determine the structure from its static property via the model of a multiquark state or a hadron molecule (in case that both kinds of  models explain well  the data of the static properties such as masses and decay widths), we can try to explore a 'production definition' via the study of the production mechanism, especially in the multi-production processes.  This means  to investigate
the production mechanism for the specific hadron to see whether it is produced like a multiquark state or like a hadron molecule \cite{Han:2009jw}. In this paper we investigate the $T_{cc}$ structure via its production mechanism.
The $T_{cc}$ state has been long studied theoretically  and has recently been observed as a narrow resonance in the $DD\pi$ spectrum by the LHCb collaboration \cite{LHCb:2021vvq,LHCb:2021auc} (for theoretical review, see \cite{Liu:2019zoy} and Refs. therein).
Its mass is just below the $D^{*+}D^0$ threshold and its quantum numbers are consistent with $I(J^P)=0(1^+)$. The static properties of this resonance have been well predicted  by both assumptions as tetraquark or molecule \cite{Wang:2017uld,Wang:2017dtg,Li:2012ss} though some potential models of multiquark state fail to give the  mass \cite{jbc}. Since the observation of this $T_{cc}$ state, discussions of double-heavy four quark states in both molecular configuration \cite{Li:2021zbw,Meng:2021jnw,Chen:2021vhg,Ling:2021bir,Feijoo:2021ppq,Yan:2021wdl,Dai:2021wxi,Wang:2021yld,Xin:2021wcr,Huang:2021urd,Fleming:2021wmk,Ren:2021dsi,Chen:2021tnn} and tetraquark picture \cite{Agaev:2021vur,Weng:2021hje,Guo:2021yws,Azizi:2021aib} are performed to explain the static and/or decay properties.
However,  $T_{cc}$ can be taken as an excellent  example for the study of the production mechanism (rather than via decay or static properties) to  gain insight on the hadronic structure, and to shed light on the confinement mechanism.  It is also special comparing to other  exotic ones like the XYZ particles, which will be explained in details later.
The analysis of this paper  serves to answer:
What is it, a multiquark bound state ($cc\bar q\bar q$) or a hadron molecule ($DD^*$)?
What have been observed and what to be observed for this purpose in the production process?
 %The static properties of this resonance have been well predicted  by both assumptions (wangzhigang, lining...)  though the potential model of multiquark state fails to give the correct mass??????
The main body of our paper emphasizes the finger prints of its production as a four-quark state.
As a comparison, we also investigate its production as a molecule by calculating the production rate, and propose methods to make judgement.
This content is in the appendix because of large uncertainties.

First of all, before paying any attention to its structure, one should notice that  this production process is characteristic of the fact that two heavy quark pairs $cc\bar{c}\bar{c}$ are produced. This is the very reason why the $T_{cc}$ is an outstanding example for the study of the production mechanism to gain insight on its structure.
The production of two heavy quark pairs is a perturbative QCD process,
%Though this is not directly related with the resonance structure, if factorization approximately valid.
    %But some subtle relation exist, even we can put up an unambiguous 'production definition' on the structure, multiquark state or molecule......}
    % More concrete, in production,
so the interface between perturbative and nonperturbative QCD plays an important r\^ole in the $T_{cc}$ production, i.e.,
if $T_{cc}$ is  a tetraquark rather than molecule of two $D$'s, the colour connection of the $cc$ quark pair   must be  a special and nontrivial one.
This colour connection  just  shares exact similarity in the production mechanism of the $\Xi_{cc}$, a baryon, which can not be produced as a hadron molecule.
  %The doubly heavy open charm   (double beauty) production,   the colour connection of the system  which  these two heavy quarks belong to are non  to produce the doubly heavy open charm hadrons.}
Therefore, once the production mechanism of $T_{cc}$ and $\Xi_{cc}$ is proved to be the same by various observables,
one can determine that $T_{cc}$ is produced as $\Xi_{cc}$ at the quark level rather than at the hadron level.
 %
 %If it is not a molecule,  \textcolor{red}{then  }  which can be used as judgement to determine whetehr it is tetra.
 This is the key point of this paper.  As a mater of fact,
  $T_{cc}$ has been investigated from this viewpoint and  suggested to be measured in various processes, e.g., $e^+e^-$ annihilation in various energies \cite{Jin:2013bra,Jin:2014nva,Jin:2015mla}, $\Upsilon$ decay \cite{Li:2020ggh}, as well as hadronic collisions \cite{Qin:2020zlg}.   Besides the production mechanism, some models on static property of the $T_{cc}$ also is fixed by the experimental data of $\Xi_{cc}$ \cite{Karliner:2017qjm}. %\textcolor{red}{
The recent discovery of the resonance is from the $pp$ collision at LHC and agrees well with the suggestion by Qin, Shen and Yu \cite{Qin:2020zlg}, in which a full analogy
to $\Xi_{cc}$ is employed as the key input for the prediction. This strongly motivates the investigation of this paper.  %}
In the next section (Sec. 2), we investigate the special colour connection for $\Xi_{cc}$ and  $T_{cc}$ production. Then in section 3 we study the hadronization and demonstrate the finger print observables for $T_{cc}$ as four-quark state. The conclusion is in section 4.

 %And for the production mechnism, it ahs been pointed out ......analous to   With  this  one can obtain many specific properties.
 %Further more thes property can be employed as a tool to distiguish it from a molecule states......

 %\textcolor{red}{After the analysis on the production mechanism, we will introduce a new 'correlation' between events, i.e., Xi cc T cc correlation ......
 %this will be compared with the DD* correlation in same event, as to distinguish the difference of production because of difference in stucture......} \cite{LHCb:2021vvq,LHCb:2021auc}

\section{Colour and baryon number fluctuation of preconfinement system and   doubly open charm hadron production}

As is mentioned above, if the $T_{cc}$ is produced as a four-quark state, there must be a $cc$ cluster produced from the primary hard interaction.
In high energy processes such as proton-proton collisions at the LHC, the hard scattering produces the $cc$ cluster, or more explicitly the $cc \bar c \bar c$,
 which transfer into possible preconfinement systems before the final hadronization process,
in  which  $T_{cc}$ or other handrons are formed. Those preconfinement systems with a specific
colour connection  favour the baryon ($\Xi_{cc}$) and $T_{cc}$ production.
The corresponding phenomena in the production will play the r\^ole of the finger prints of the four-quark state structure of $T_{cc}$.
 %as thenfluctuation of the baryon numbe r of the clusters.

 %\subsection{preconfinement}

The concept of preconfinement is introduced by  Amati and Veneziano \cite{AV}.  In our case,
the production of the charm quark system $cc \bar c \bar c$ can be calculated by pertubative QCD.
In general, it  is accompanied by gluon radiation and light quark pair production. The dominant contribution can be described
in a parton evolution picture, which  is embedded in, e.g.,  parton shower  models. %So this process is described as hard ME+PS.  At last a
The preconfined system is considered as the end of this evolution, which is typically of the formation of colour-singlet clusters with modest masses
independent on the center-of-mass energy of the scattering process. Each colour-singlet cluster will independently hadronize into some hadrons.
The preconfinement cluster formation has many evidences such as the local parton hadron duality (for details, see \cite{Li:2019vrc} and refs. therein).

 %In general, it is ...... colour neutral flow

It has long been argued that the colour structure of the preconfined system is not unique \cite{our2,gsj}.
%We will in the following discuss a specific one that leads to the $T_{cc}$ and $\Xi_{cc}$ production.
It has also been recognized that different colour structures will lead to different non-trivial baryon number distribution of the colour-singlet clusters,
which is referred as the baryon number fluctuation of the preconfined system \cite{Li:2019vrc}.
Some phenomena such as the baryon number enhancement (with saturation) has been investigated in \cite{Li:2019vrc},
which is also an important feature of the $T_{cc}$ production as a four-quark state.
These issues of the preconfinement can be easily understood from the
analysis  on the colour connection of the most simple colour-singlet four quark system $ c\bar{c} c\bar{c}$ produced from some perturbative  processes (e.g.,   \cite{Jin:2013bra}).  % which    has the following  property:

The $cc\bar{c}\bar{c}$ colour space  can be decomposed as Eqs. (1) and (2)  in the introduction section or in other ways, but we investigate the specific one
\begin{equation}
(3_{1}\otimes3_{2})\otimes(3_{1}^{*}\otimes3_{2}^{*})=(3_{12}^{*}\oplus6_{12})\otimes(3_{12}\oplus6_{12}^{*})=(3_{12}^{*}\otimes3_{12})\oplus(6_{12}\otimes6_{12}^{*})\oplus\cdots,\label{diqcl}
\end{equation}
where $3_{12}$ ($3_{12}^{*}$) denotes the triplet (anti-triplet) representation of the $SU_{c}(3)$ group and
$6_{12}$ ($6_{12}^{*}$) denotes the sextet (anti-sextet) representation of the $SU_{c}(3)$ group. When two (anti)quarks in the
color state $3^{*}(3)$ attract each other  %and form a(n) `(anti)diquark'
and their invariant mass is sufficiently small, such a cluster %has a certain probability
 can be considered as  a(n) (anti)diquark \cite{Jin:2013bra}.
%by triggering the leading baryons and two-jet-like event shape.
   %
 For the case in which only one of the pair has a small invariant mass, the color configuration can be better written as
\begin{eqnarray}
(3_{1}\otimes3_{2})\otimes3_{1}^{*}\otimes3_{2}^{*}=3_{12}^{*}\otimes3_{1}^{*}\otimes3_{2}^{*}\oplus\cdots, & \text{or}\nonumber \\
3_{1}\otimes3_{2}\otimes(3_{1}^{*}\otimes3_{2}^{*})=3_{1}\otimes3_{2}\otimes3_{12}\oplus\cdots.
\label{q2qq}
\end{eqnarray}
The above equation  % should be taken into account in considering the hadronization model of the process $e^+e^- \to c\bar{c} c\bar{c} \to \Xi_{cc}^+ + X$.
indicates that after the perturbative QCD evolution for the whole system, it is possible to form clusters, like 'ccq + (other quark anti-quark pairs) + gluons', with cc  in the 3*.
This is a colour singlet  cluster with a non-zero baryon number in the preconfinement system.
It can enhance baryon production, especially for the $\Xi_{cc}$, as displayed in Fig 1.
%(Moriond talk, baryon production, saturation).
On the other hand, one can see from Fig. 2 that this colour structure can also lead to  $T_{cc}$ production, if $T_{cc}$ is indeed a four-quark state.
After the $T_{cc}$  formed, there can be a remnant diquark, which again can enhance the baryon production as Fig. 1.

From the above analysis, such a specific colour and baryon number fluctuation of the preconfinement system enhances both $\Xi_{cc}$ and $T_{cc}$.
Et vice verse,  the doubly charm baryon or tetraquark play the role of the 'trigger' to probe this  special colour connection.
In brief, the $T_{cc}$ and $\Xi_{cc}$ production requires the above physical mechanism happen on the interface between PQCD and NPQCD.
Therefore, the corresponding unique signals can be taken as the fingerprints for judging that the resonance structure is a true tetraquark
rather than a molecule. %See next subsection for details.
The details of the  hadronization and the signals are investigated in the next section.

\section{Hadronization of cc diquark  and finger prints of tetraquark in experiment}

  %e.g., spectrum wrt Xicc, dip in theta distributuion for both Xi cc as well as T cc....

The preconfinement  and subsequent hadronization are `branching' processes via the creation of quarks from the vacuum by the strong interactions within the system.
The created quarks and the primary quarks are combined into color-singlet clusters and then hadrons at last.
In Eq. (\ref{q2qq}), the colour configuration as a whole is like a `big antibaryon' ($3^*3^*3^*$) but with a large invariant mass. For example, in Figs. 1, 2,
The diquark $cc$ must combine with a quark $q$ (or antidiquark) to form a  color-singlet system, $\Xi_{cc}$/$T_{cc}$.
To balance the quantum numbers of color and flavour, an antiquark/diquark must be created simultaneously.
To branch them further, more quark pairs and diquark pairs will be created from the vacuum via
the interactions among the quark system. Such a cascade process
will proceed until the end of time, when most of the `inner energy'
of the entire system is transformed into the kinematical energies
and masses of the produced hadrons. In this process some clusters with non-zero baryon number produce because of some diquark pair creation.
Each of the two primary $\bar{c}$'s combines with a created quark or antidiquark to respectively hadronize into two open charmed hadrons.
To quantitively describe such a nonperturbative hadronization procedure, we adopt a concrete hadronization model, the Lund string model \cite{string},
which is realized by PYTHIA/JETSET \cite{pythia}.
For the configurations considered here, the above process is
straightforward, except that for each step, we must assign special
quantum numbers for each specific kind of hadrons according to their production rates.

\begin{figure}[htb]
\centering
\scalebox{0.35}{\includegraphics{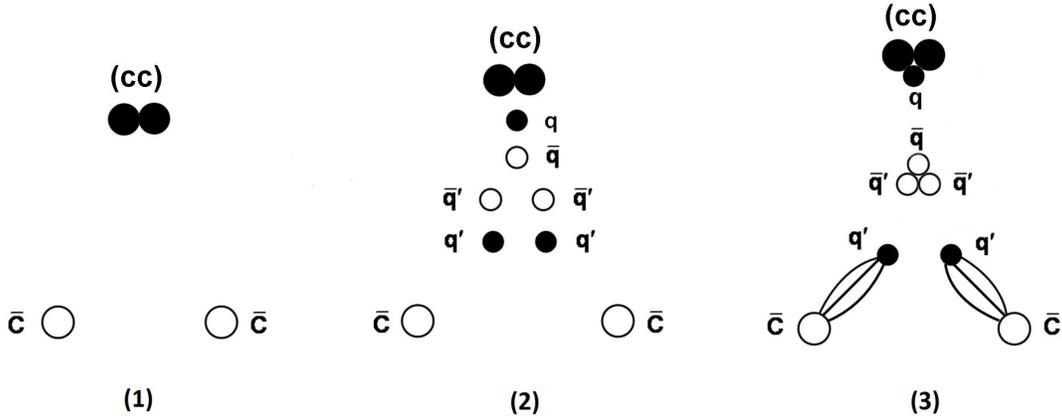}}
\caption{The $\Xi_{cc}$ production and the string formation for the hadronization of the $(cc)\bar{c}\bar{c}$ system with the aid of quark creation from the vacuum. Solid circles represent quarks, while hollow circles represent antiquarks.
The primary $\bar c \bar c$ connect to quarks respectively via two strings.}
\label{cc2xi}
\end{figure}

\begin{figure}[htb]
\centering
\scalebox{0.21}{\includegraphics{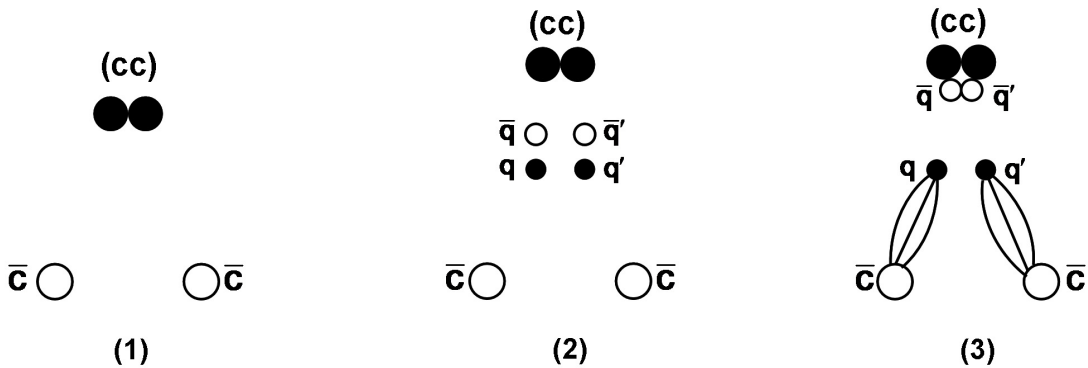}}
\caption{The $T_{cc}$ production and the string formation for the hadronization of the $(cc)\bar{c}\bar{c}$ system with the aid of quark creation from the vacuum. Solid circles represent quarks, while hollow circles represent antiquarks.
The primary $\bar c \bar c$ connect to quarks respectively via two strings.}
\label{cc2tcc}
\end{figure}

In the case of $\Xi_{cc}$ production, the complementary antiquark can produce
an antibaryon by combining with an antidiquark, and then the
balancing diquark is broken by the interactions with the remaining system and then each becomes connected to the two primary $\bar{c}$'s to form
two strings. The resultant hadronization can be described by the conventional string-fragmentation picture.
This procedure is illustrated in FIG.\ref{cc2xi}.
In the case of $T_{cc}$ production, if the complementary diquark is broken
and proceeds in the same manner described above, the procedure can be described by FIG.\ref{cc2tcc}.
However, it is also possible that the new created diquark is not broken, and then the hadronization proceeds like the $\Xi_{cc}$ production shown in Fig.\ref{cc2xi}.
In this scenario, the baryon production is  also enhanced.
Therefore, there is baryon enhancement in both the $\Xi_{cc}$ and $T_{cc}$ production, and this similarity is expected to be observed.
We hereby propose an observable to measure this effect. First find a cluster including a $\Xi_{cc}$ or $T_{cc}$ hadron and other particles with relative small momenta,
and then measure the invariant mass of this cluster $M_{T(\Xi)}+ \Delta $ ($M_{T(\Xi)}$ is mass of $T_{cc}(\Xi_{cc})$) and the total baryon number $N_B$ (algebric sum or arithmetic sum). The $N_B$ as a function of $\Delta$ is thus obtained.  The value of $N_B$ should be  significantly larger than the averaged baryon production rate in various high energy scattering processes, which is around 10 per cent with respect to all hadrons.

%({\bf QQ:} It is not clear to me how to judge the baryon enhancement even with this observable measured.
%Should it be compared to some other clusters, e.g. a $D$ meson cluster or $X(3872)$ cluster?)}

\begin{figure}[htb]
\centering
\scalebox{0.96}{\includegraphics{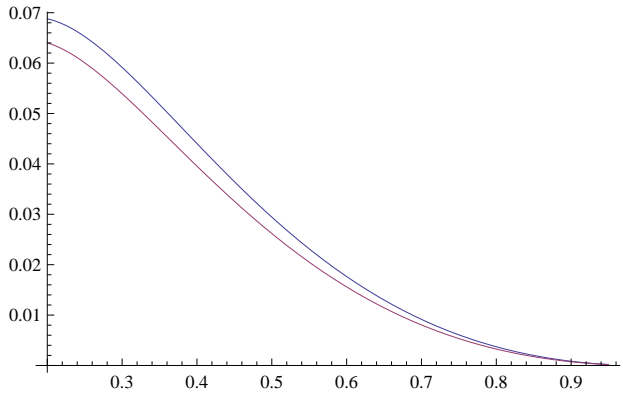}}
\caption{Comparison of the $z$-distributions of the fragmentations (from the $cc$ diquark) to $\Xi_{cc}$ (blue) and $T_{cc}$ (red).}
\label{pus}
\end{figure}
The next observable that we propose to measure on the similarity between the $\Xi_{cc}$ and $T_{cc}$ is  their kinematic spectra.
The fragmentation of the heavy diquark can be described by the Peterson formula \cite{pet83}
\begin{equation}
f(z)\propto\frac{1}{z(1-1/z-\epsilon_{Q}/(1-z))^{2}} \; ,\label{Peterson}
\end{equation}
where $z$ is defined by $p_+^\text{hadron}/p_+^\text{cluster}$ with $p_+$ being the sum of the energy and the momentum projected in the
moving direction of the cluster, and the free parameter $\epsilon_{Q}$ is expected to scale between
flavours as $\epsilon_{Q}\propto1/m_{Q}^{2}$. In practice, different $\epsilon_{Q}$ values inversely proportional to
the hadron masses squares are used for $T_{cc}$ and $\Xi_{cc}$. The results for $f(z)$ of $T_{cc}$ and $\Xi_{cc}$ are displayed in
FIG.~\ref{pus} to show the difference.
It can be seen that for reasonable kinematics region, the differences between these two hadron can be neglected.
Therefore, this similarity in the line shape of the momentum spectra with respect to those of $\Xi_{cc}$ is another  feature to judge $T_{cc}$ as a four-quark state, which can be examined in experiments.

\begin{figure}[htb]
\centering
\scalebox{0.4}{\includegraphics{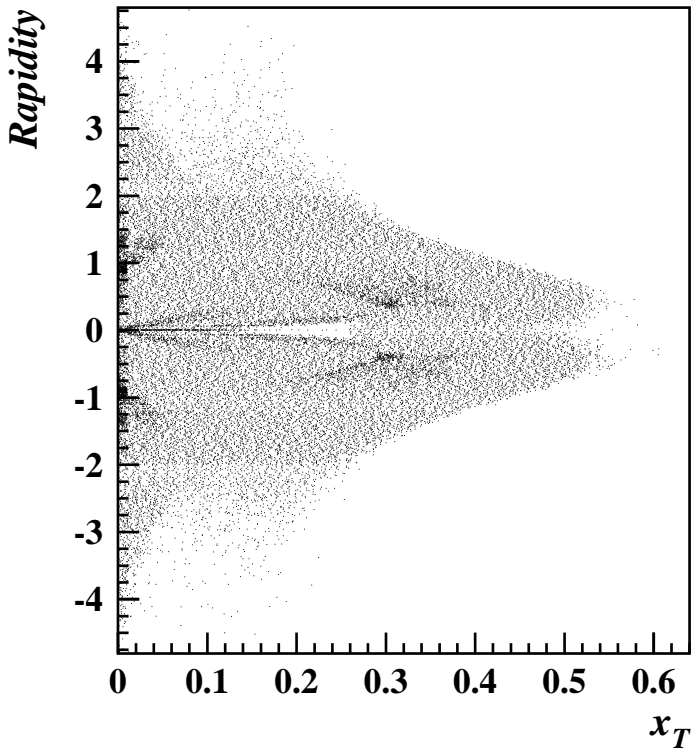}}
\caption{2D distribution onrapidity and $x_T=p_T/\sqrt{s*}$.}
\label{pu}
\end{figure}

To further prove this physical picture, we propose that the light hadrons produced adjointly with $\Xi_{cc}$/$T_{cc}$ can provide additional clues.
One notices that the two-string structure in FIG.~\ref{cc2xi}\&\ref{cc2tcc} is typical for this kind of production processes.
The reason is that however the $cc$ diquark fragments, to balance the colour, the subsequent  hadronization of the remnant except the $cc$ diquark cluster can only happen
 in two (or more)  colour-singlet clusters rather than only one.
Thus, one can measure this special property, the string effect, which was suggested by the Lund group in 1980s \cite{string}.
In experiments, one should get quite similar kinematic distributions such as those of the energy, momentum, etc., for $\Xi_{cc}$ and $T_{cc}$.
To visualize this effect, a simulation is performed.
The fragmentation of the complementary (anti) quark is handled by the fragmentation function employed by the
LUND group \cite{string}
\begin{equation}
f(z)\propto z^{-1}(1-z)^{a}exp(-bm_{\perp}^{2}/z),\label{Anderson}
\end{equation}
where $a$ and $b$ are free parameters. In our simulation program, we take $a=0.3$ GeV$^{-2}$
and $b=0.58$ GeV$^{-2}$, as used in PYTHIA~\cite{pythia}.
The fragmentation of the strings can be referred to the classical book
about the Lund model by B. Anderson~\cite{string} and the PYTHIA manual~\cite{pythia}.
With the simulated events, we first have to reasonably define a cluster, whose total transverse momentum is zero at the jet level.
In practice, we approximately use the doubly heavy hadron momentum as jet momentum, i.e., the Peterson function taken as a $\delta$ function.
In this scenario, we collect hadrons around it to balance the transverse momentum of  $\Xi_{cc}$/$T_{cc}$ to obtain the cluster of hadrons (with invariant mass $\sqrt{s*}$) to be investigated. Then, we should Lorentz transform this cluster to its rest frame.
Take the direction opposite to the $\Xi_{cc}$/$T_{cc}$ momentum as the $z$ axis direction, then we calculate the rapidities and re-scaled transverse momenta $x_T\equiv p_T/\sqrt{s*}$ of the  hadrons in the two string system to get the distribution, which is shown in FIG.~\ref{pu}.

The above suggested  'similarity' study is very plausible since the experiment \cite{LHCb:2021auc} has attempted to
make comparisons between the $T_{cc}^+$ signal and the $DD$ pair production,
which is expected to be contributed significantly from double scattering.
Nevertheless, more data are required.
Furthermore, one should give more definite confirmation on the  the new discovered resonance, as well as on  $\Xi_{cc}$.
To eliminate some kinematic uncertainty, future measurements  by other kinematic region,  such as the CMS collaboration in the central rapidity region, are valuable.

Except for the similarity between the kinematics distributions of the $\Xi_{cc}$ and $T_{cc}$ production processes, their total production rates can also
be a helpful observable to distinguish the nature of $T_{cc}$. If $T_{cc}$ is a four-quark state, both
$\Xi_{cc}$ and $T_{cc}$ are generated from the fragmentation of the heavy diquark $cc$ and are managed by Eq.~\eqref{Peterson}.
In this scenario, Qin, Shen and Yu \cite{Qin:2020zlg} estimated the production ratio of $T_{cc}$ which was assumed to be a isospin singlet state.
The key point is that, taking into account all these similarities, they gave a reasonable conjecture of the production ratio, $T_{cc}$/ $\Xi_{cc}$ $\sim$ 1/4.
It is quite significant that with this estimation, they have predicted the signal yield which agrees well with the data \cite{LHCb:2021vvq,LHCb:2021auc}.
It was also estimated in \cite{lsy2016} that the diquark versus quark fragmentation ratio as $\lambda/2$ ($\lambda \sim 0.3$), which is a result without taking  the isospin counting.  The present observed one is composed with the $ud$ antidiquark, and it is thus 2 times of $uu$ or $dd$ case. Consequently, an estimation around the range 1/3-1/5 is generally reasonable and adoptable, consistent with \cite{Qin:2020zlg}.
In contrast, if $T_{cc}$ is regarded as a molecule state, the production rate would be smaller by one order of magnitude, which
is disfavored by the current data \cite{LHCb:2021vvq,LHCb:2021auc}. The details can be found in Appendix \ref{sec:app}.

In the end, it is valuable to examine whether there exists a further resonance,
here denoting as $T'_{cc}$, which has a different isospin from $T_{cc}$. The relative production rate of $T_{cc}/T'_{cc}$ is
a good observable to decide whether $T_{cc}$ has the four-quark state structure.
The reason is as follows. Consider the $cc$ as a $3^*$ anti heavy 'quark', then $T_{cc}$ and $T'_{cc}$ are anomalous to heavy anti-baryons from the colour aspect.
By such a viewpoint, their production ratio follows the mass and production rate rules observed from the  experiment \cite{Belle:2017caf} for heavy baryons,
dependent on the isopspin structures.
More explicitly, the mass splitting and production rate between $T_{cc}$ and $T'_{cc}$ should be similar to $\Lambda_c$ and $\Sigma_c$,
with the production rate of $\Sigma_c$ is smaller than that of $\Lambda_c$ by one order of magnitude.
If the present observed $T_{cc}$ is isospin singlet and the $\Lambda_c$-like one, then a scan to the larger mass region with more data by LHCb can
find a isospin-1 resonance $T'_{cc}$ with a lower production rate. If, on the contrary,  the observed $T_{cc}$ is the $\Sigma_c$-like one, then a scan to
the smaller mass region will find such a isospin-0 $T'_{cc}$ with a much larger production rate.
Now the experiment \cite{LHCb:2021auc} favours a $\Lambda_c$-like $T_{cc}$, since there is not another significant resonance observed
and the signal yield is consistent with the estimation of \cite{Qin:2020zlg}.
Based one the above discussion, the production rate is a perfect evidence for the four-quark state production mechanism.

\section{Conclusion}

The consistence between the theoretical analysis on the $T_{cc}$ production by Qin, Shen and Yu \cite{Qin:2020zlg}
and the data \cite{LHCb:2021vvq,LHCb:2021auc} strongly favours the newly discovered resonance $T_{cc}$ is produced as
a real four-quark state.
We in this paper clarify the production mechanism, and provide further experimental observables to look for the finger prints of the $T_{cc}$ as a tetraquark.
The observables include:
the similarity of $T_{cc}$ and $\Xi_{cc}$ in the kinematics spectra, in the baryon number enhancement, in the string effect, etc.
Moreover, measurements of the production cross section of $T_{cc}$ and of the production ratio between $T_{cc}$ to a different
isospin state $T'_{cc}$ (if it exists) would also be very helpful.
With these measurements, we will have a good chance to confirm or deny the $T_{cc}$ produced as a four-quark state.

\acknowledgments

We thank Prof. Wang Zhi-Gang for the discussion on application of QCD sum rule in predicting the mass of $T_{cc}$,  Prof. Jiang Jun  for the discussion in some aspects of the hadron molecule and Prof. Li Ning  (Sun Yat-Sen Univ.) for the nonrelativistic wave functions.
This work is supported in part by  National Natural Science Foundation of China (grant Nos.  11635009, 11775130, 11775132, 11975112 and 12005068).

%%%%%%%%%%%%%%%%%%%%%%%%%%%%%%%%%%%%%%%%%%%%%*******************************
\appendix \section{$T_{cc}$ production as  DD* molecule}\label{sec:app}

In this appendix, as mentioned in the  introduction, when  we take '$T_{cc}$' as the hadron molecule {\it in production}, we mean that such a production process includes two factors if we can have a factorized formulation: First is the productions of D and D* hadrons; Second is the
description on how these two hadrons D and D* combined to a 'hadron molecule', which, if allowed, we also call '$T_{cc}$'.

For the second factor, there could be two possibilities. Corresponding to these two ways, the requirement of the information  to the first factor is somehow also different.  These two factorized formulations are investigated in the following.

\subsection{Non relativistic wave function formulation \cite{Jin:2016vjn,lsy2016,r2005lsy} and  DD* momentum correlation measurement as smoking gun}

If the molecule composed of DD* can be described by a non-relativistic wave function (the $T_{cc}$ mass also got by solving the corresponding Schr\"odinger function), then we can rely on this wave function to  study
the production of the molecule of DD*, with the DD* produced from the hard interaction in the  collision, such as pp scattering at LHC. At last one gets a factorized formulation.

\begin{figure}
	\centering
	\vspace*{-1.5cm}
	\hspace*{-1.5cm}\includegraphics[scale=1.0]{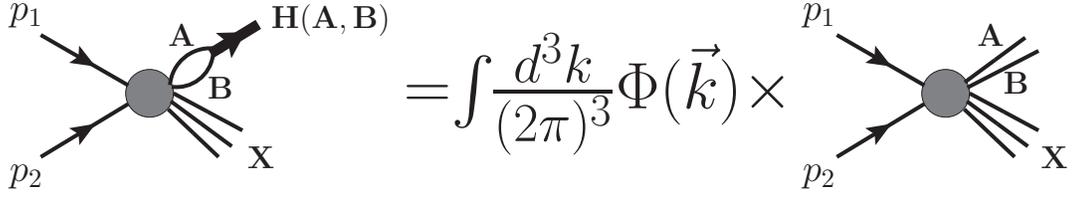}
	\vspace*{-4.0cm}
	\caption{Any hadron molecule  $H(A,B)$ production  process $p(p_1)p(p_2) \to A(p_A)+B(p_B) + X \to H(A,B)(P_H) +X$.
		\label{feygl}}
\end{figure}

 The process $pp \to D+D* + X \to T(DD*) +X$ is illustrated in Fig.~\ref{feygl}, and the corresponding invariant amplitude is:
\begin{equation}
	\label{startpoint}
\begin{array}{l l l}
A_{inv} & = & \langle T(DD*), X|\hat{T}|p p \rangle  \\
     & = & \frac{1}{\sqrt{\frac{{{m}_{D}}{{m}_{D*}}}{{{m}_{D}}+{{m}_{D*}}}}} \int \frac{d^3 k}{(2\pi)^3} \Phi(\vec k)
           \langle D|\langle D*|\langle X|\hat{T}|pp\rangle  \\
     & = & \frac{1}{\sqrt{\frac{{{m}_{D}}{{m}_{D*}}}{{{m}_{D}}+{{m}_{D*}}}}} \int \frac{d^3 k}{(2\pi)^3} \Phi(\vec k)
           {\bold M}(\vec k).
%\end{eqnarray}
\end{array}
\end{equation}
This formulation is valid in the rest frame of $T(DD*)$, the bound state of the ingredient hadrons $D$ and $D*$. In the above equation, $\vec k$ is the relative 3-momentum between $D$ and $D*$ in the  rest frame of $T(DD*)$. ${\bold M}(\vec k)$ is the invariant amplitude  for the free (unbound) $D$ and $D*$ production. The factor $\frac{1}{\sqrt{\frac{{{m}_{D}}{{m}_{D*}}}{{{m}_{D}}+{{m}_{D*}}}}}$ comes from the normalization of the bound state to be $ 2 E_H V$ just as a single particle.   The normalization for the  wave function $\Phi(\vec k)$ in momentum space (Fourier transformation from the configuration space) is
\begin{equation}
     \int \frac{d^3 k}{(2\pi)^3} |\Phi(\vec k)|^2 =1.
\end{equation} From the above equation, it is obvious that if we know the analytical form of the wave function as well as the free particle invariant amplitude, we can calculate the amplitude of the bound state simply by integrating the relative momentum $\vec k$.
In practice, certain decomposition and simplification  will be taken for a concrete $^{2S+1}L_J$ state. The resulting formulations are covariant.

To get a factorized formulation, we should expand the $\Phi(\vec k)$ around $\vec k=0$. This is especially reasonable in the heavy hadron molecule case since the typical relative momentum    $\vec k$ is rather small compared to the mass of the molecule. The binding energy is generally very small. %(may say more about some molecule models).
For two orbital angular momentum states L=0, 1,
taking into account the parity of these two wave functions,
one respectively obtains
$$\int \frac{d^3 k}{(2\pi)^3} \Phi(\vec k) {\bold M}(\vec k) =\Psi(0){\bold M}(\vec k)|_{\vec{k}=0}~~~(L=0), $$
$$ \int \frac{d^3 k}{(2\pi)^3} \Phi(\vec k) {\bold M}_\lambda(\vec k) =\Psi'(0)\epsilon_\lambda ^\mu \partial^k _\mu{\bold M}(\vec k)|_{\vec{k}=0}~~~(L=1),$$
to the leading contribution.
Here $\Psi(0)$ or $\Psi'(0)$ is  the configuration space wave  function or divergence of the configuration space wave  function at origin. $\epsilon_\lambda^\mu$ is the polarization vector of the vector hadron D*.

Since $T_{cc}$ is considered as $J^P=1^+$, one can assume the dominant contribution is the S wave.
In this case the production rate or cross section can be calculated by these two factors: The absolute square of the  wave function at origin, and the distribution (differential cross section) with respect to the relative momentum  $\vec k$ extrapolated to $\vec k=0$.

The production rate of the P wave particles  will be quite small as argued by us in \cite{Jin:2016vjn} because $\partial^k _\mu\bold M(\vec k)|_{\vec{k}=0}$ is much more suppressed. The reason for this is that $\bold M(\vec k)$ is slowly varying on $\vec k$. This can be checked by numerical study on the correlations \cite{Jin:2016vjn}. The following numerical result can show such a fact. In general, because of the property of the  spherical harmonics,  typical of the   the non-relativistic (assuming central field) wave function, the  L order partial wave is relevant to the L order of  moment of the function which convolute with the L order spherical harmonics.
So the L order of the derivative of the amplitude enters.  For slowly varying behaviour of the D D* correlation, the high order of L partial wave contributions are generally suppressed in production for this model.
Needless to say, the high order L states/partial wave components could be a good probe to the details of the correlation in the  production.
 %This can be used to
%probe the L structure of the molecule.

The details of the molecule structure,  e.g., the isospin effects,  is relevant to the absolute production rate. The reason is that different isospin structure predicts different distribution near the  origin.   The weaker bound, the smaller value near the origin, and the lower production rate predicted.
In \cite{Li:2012ss}, it is predicted that only I=0 bound state exists. This indicates a qualitatively direct judgement, i.e.,  to measure the isospin of the resonance, to see whether isospin multi-states  exist \cite{LHCb:2021auc}.

To study the production of the hadron molecule, the other key point is to study the relevant  two hadron  correlation in momentum space.
The free pair cross section $p(p_1)p(p_2) \to A(p_A)+B(p_B) + X $ can be expressed as:
\begin{eqnarray}
	\label{int2}
	&& \frac{1}{N} \frac{d N}{d^3 P_H d^3 q} \propto \frac{1}{F} \sum_{j \neq A, B} \hspace{-0.67 cm} \int \prod \frac{d^3 p_j}{(2 \pi)^3 2 E_j} \nonumber \\
	&& \times \overline{ |\hat{O} |^2} (p_j, P_H=p_A+p_B, q=p_A-p_B)  \\
	&& \times (2 \pi)^4 \delta^{(4)}(P_{intial}-\sum_{j \neq A, B} p_j-p_A-p_B). \nonumber
\end{eqnarray}
Here the average is on various spin states, and the proper initial flux factor $1/F$ and phase space integral are needed. $\hat{O}$ is the amplitude of production of two free ingredient particles (with vanishing relative momentum and proper angular momentum state). It is not possible to be calculated directly with some effective quantum field theory/model when the initial state is (anti) protons and $A$ and $B$ are hadrons or clusters. However, it can be obtained with an event generator such as PYTHIA \cite{pythia} or equivalently Shandong Quark Combination Model \cite{jin2010} etc., for the case that $A$ and $B$ are both on shell.
% It is the advantage that in the above formulation only the on shell case is considered.
It is the advantage that in the nonrelativistic  framework we employ, only the on shell case is considered, so that the numerical calculation with event generator is plausible.
The quantity of Eq. (\ref{int2}) describes the two hadrons/clusters ($A$ and $B$) correlation in the phase space. For the hadron case, by proper integral on components of $P_H$ and/or $q$, the resulting correlations can be directly compared with
data and serve for tuning the parameters. This strongly implies that to study various hadron molecule, careful measurement on the two (corresponding) hadron (momentum) correlation in the corresponding scattering process is crucial, though such study is in fact quite lack  until now.

Since the special physical picture of the non-relativistic framework, it is only valid in the rest frame of the two ingredient particles. One can define the following covariant space-like relative  momentum $\hat{q}$ as
\begin{equation}
	\hat{q}=(p_A-p_B) - \frac{(p_A-p_B) \cdot (p_A+p_B)}{(p_A+p_B)^2}(p_A+p_B).
\end{equation}
It is clear that in the rest frame of $A$ and $B$ ($H(A,B)$) where $\vec{p}_A+\vec{p}_B=0$, $\hat{q}=(0,\vec{k})$ and the $k=\sqrt{-\hat{q}^2}$ is exactly the absolute value of the 3-relative momentum $|\vec{p}_A-\vec{p}_B|$.

Employing the event generator, one gets
\begin{equation} \label{ext}
	\frac{1}{N} \frac{d N}{d^3 P_H d^3 \hat{q}}, \forall \hat{q},
\end{equation}
then extrapolates to the special case $k=0$. Numerically, one can take an average around $k=0$ for the above quantity.
Then we get, up to the kinematic factors as for the covariant form,
\begin{eqnarray}
	\label{int}
	&& \frac{1}{F} \sum_{j \neq A, B} \hspace{-0.67 cm} \int \prod \frac{d^3 p_j}{(2 \pi)^3 2 E_j} \overline{ |\hat{O} |^2} (p_j, P_H=p_A+p_B, k=0) \nonumber \\
	&& \times (2 \pi)^4 \delta^{(4)}(P_{intial}-\sum_{j \neq A, B} p_j-P_H).
\end{eqnarray}
%Eq. (\ref{int}) is exactly the differential cross section of the bound state $H(A,B)$ divided by $|\Psi(0)|^2$.

There are several very basic facts supporting the extrapolation.
First of all, the amplitude and cross section are  analytical in phase space. Any practical generator should reproduce this property, and any ultraviolet divergence is not present. Secondly, the study of strong interaction is complex because of the SU(3) non-Abelian interaction, but its simulation has one simplicity: All particles taking part in the strong interactions are massive, which eliminates the infrared singularities.

%Some numerical.......

\begin{figure}
\label{ckc}
	\centering
	\scalebox{0.20}{\includegraphics{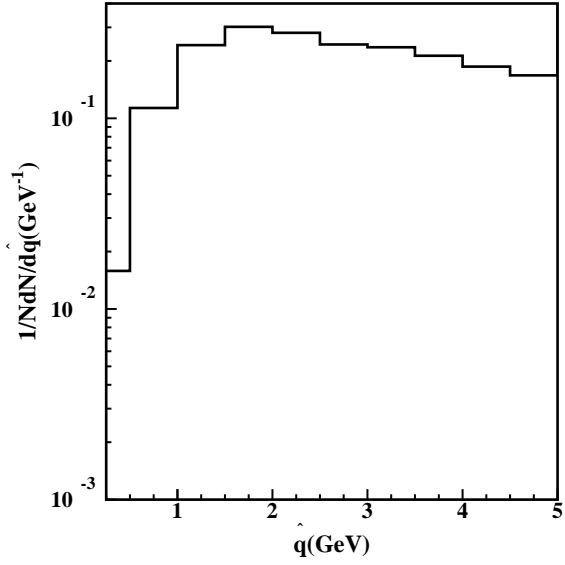}}
	\caption{The distribution of 3-relative momentum $k$ between DD* calculated by PYTHIA.}
	\label{ptbs}
\end{figure}

From  Figure \ref{ckc} we can see, the distribution on the relative momentum between D and D* is quite flat. The correlation quite near k=0 is generally not reliable in   generators because of lack of data to test. In general we can rather rely on the average around the small value. However, a careful  measurement on this correlation surely plays the r\^ole of  the smoking gun, i.e., if from experiment we make sure this distribution on  $\vec k$ is vanishing on the value  $\vec k=0$, such molecule will not possible to be produce from the corresponding scattering process. As an illustration of the method rather than the final conclusion,   we estimate the production rate and cross section for a S wave isospin-singlet DD* molecule \cite{Li:2012ss} based on  the above calculation.  We use  the wave function at origin value as ($0.1407 GeV^{3/2}$ )\cite{Li:2012ss} and  the cross section of $pp \to c \bar c c \bar c +X$ as $3.6 \times  10^7$ pb by NLO calculation.  The  cross section $pp \to T_{DD^*} +X $  is around $3 \times 10^{2} pb$, which   is one order lower than that of  the production rate of the  four-quark state \cite{Qin:2020zlg}. % up to now we only consider this result

\subsection{Combination matrix elelment formulation \cite{Li:2005hh,r2021lsy}}

In high energies like LHC, the multi-scatterings and  underlying events are significant. It is possible that these two D's are from different scattering processes, or  that one D/D* produced from the hard scattering flying by and to combine with a D*/D from the underlying events. The Experiment has studied  some similarity with the DD pair with small invariant mass which is expected significantly contributed from double scattering \cite{LHCb:2021auc}.
To calculate the combination of a D/D* with the underlying D*/D, or with D*/D from another scattering, we again need a factorized formulation.
By the analogy to the quark level combination \cite{Li:2005hh,r2021lsy}, we can conjecture that the  cross section    to produce a molecule $T_{cc}$ with four momentum $(E, \vec K)$ in pp scattering can be written as
  \begin{equation}
\label{cossnum}
%\begin{array}{ll}
2 E \frac{d {\sigma_C}}{d^3 K}= \sum \limits_{ab} \int{dx_1 dx_2}  f^a_1(x_1)  f^b_2(x_2)
\frac{d \hat{\sigma}_{ab}}{d {\cal I}} \frac{1}{\xi^2} \frac{(2\pi)^2}{(2M)^2}  P(\xi_l) \tilde{F}(\xi, \xi_l)| _{\xi+\xi_l=1}.
%\end{array}
 \end{equation}
 %
 %%\footnote{We notice that $E \frac{d \sigma}{d^3 K}$ is just $\frac{1}{2 \pi} \frac{d^2 \sigma}{d y K_\perp dK_\perp}$, which is the observable measured at RHIC\cite{taian} }
Here $f^a_A (x_1)$ and $f^b_B(x_2)$ are the initial parton distributions with momentum factions $x_1$, $x_2$. $ M$ is the mass of the $T_{cc}$.  $P(\xi_l)$ is  the distribution of the D/D* in underlying events or the softer one from another ascattering and $\tilde{F}(\xi, \xi_l)| _{\xi+\xi_l=1}$ is the combination matrix element.
$d{\cal I}$ is the dimensionless invariant phase space for the `2-body'  final state  $D_A + X$ (Here $D_A$ refers one of the D/D* produced from the hard interaction) where $X$  treated as   one particle, with $\xi=E_{D_A}/E$.
This formula is also correct for higher order cross sections.
%This is the start point of the following numerical estimation.

For this production mechanism, the $T_{cc}$ may behave like a single D/D* at large transverse momentum.
Since both the combination matrix element as well as the underlying event distribution are yet  unknown,  we will not discuss the numerical result of this combination process for hadron molecule production.

%2) numenrical
 %DD* correlation

\end{document}